\documentclass[aps,prl,twocolumn,groupedaddress,amsmath,amssymb]{revtex4-2}

\usepackage{graphicx}
\usepackage{xcolor}
\usepackage{verbatim}
\usepackage[colorlinks=true,urlcolor=blue,linkcolor=blue,citecolor=red]{hyperref}

\newcommand{\dd}{\mathrm{d}}
\newcommand{\bnabla}{\boldsymbol{\nabla}}
\newcommand{\phiS}{\phi_\mathrm{S}}
\newcommand{\phig}{\phi_\mathrm{g}}
\newcommand{\phiR}{\phi_\mathrm{R}}
\newcommand{\chisp}{\chi_\mathrm{sp}}
\newcommand{\Eqref}[1]{Eq.\nobreakspace\eqref{#1}}
\newcommand{\Figref}[1]{Fig.\nobreakspace\ref{#1}}

\newcommand{\mysection}[1]{}
\newcommand{\mysubsection}[1]{\emph{#1}---}

\newcommand{\kBT}{{k_{\mathrm{B}}T}}
\newcommand{\boldx}{{\boldsymbol{x}}}
\newcommand{\Eqsref}[1]{Eqs.\nobreakspace\eqref{#1}}
\newcommand{\Secref}[1]{Appendix\nobreakspace\ref{#1}}
\newcommand{\mypar}[1]{\smallskip\par\noindent\emph{#1}---}

\begin{document}

\title{Phase coexistence and nucleation of aging condensates}

\author{Yicheng Qiang}
\author{Chengjie Luo}
\author{David Zwicker}
\affiliation{Max Planck Institute for Dynamics and Self-Organization, Am Fa{\ss}berg 17, 37077 G{\"o}ttingen, Germany}

\begin{abstract}
Biomolecular condensates form by phase separation and often age over time by transitioning from liquid-like droplets to solid-like states.
To capture the fundamental behavior of aging condensates, we connect the Flory–Huggins model of phase separation with the trap model of aging dynamics.
This combined model yields a rich phase diagram, including a regime of metastable coexisting liquids.
Interestingly, the dense metastable liquid serves as an intermediate state in the nucleation pathway of the glass, which thus nucleates in two steps.
Our results are generic, since they follow from a basic free-energy structure, providing a unified foundation for describing the thermodynamics and kinetics of aging biomolecular condensates.
\end{abstract}

\maketitle

Droplets composed of complex molecules, such as biomolecular condensates~\cite{Brangwynne2015,Berry2018,Choi2020a}, rarely behave as simple liquids.
In fact, many condensates slowly transition from liquid-like to solid-like rheology during \textit{in vitro} experiments~\cite{lin2015Formation,patel2015Liquid,molliex2015Phase,Murray2017a,jawerth2020Protein,Emmanouilidis2024,das2025Tunable}, a process known as \emph{aging}. %
Aging condensates are thought to play a role in neurodegenerative diseases~\cite{Alberti2021} and have been studied using detailed particle-based simulations~\cite{Harmon2017,Ranganathan2020,biswas2024Molecular}.
These studies revealed that aging can arise from several distinct microscopic mechanisms: a percolating network of reversible crosslinks can drive gelation~\cite{sciortino1993Interference,semenov1998Thermoreversible,hikmet1988Thermoreversible}, a slowing of molecular motion can drive a glass transition~\cite{testard2011Influence,sappelt1997Spinodal,berthier2011Theoretical}, or ordered aggregation can lock molecules into amyloid-like fibrils~\cite{patel2015Liquid,molliex2015Phase,Murray2017a}.
However, the general interplay between phase separation and aging remains poorly understood.

\begin{figure}[bt]
    \centering
    \includegraphics[width=1.0\linewidth]{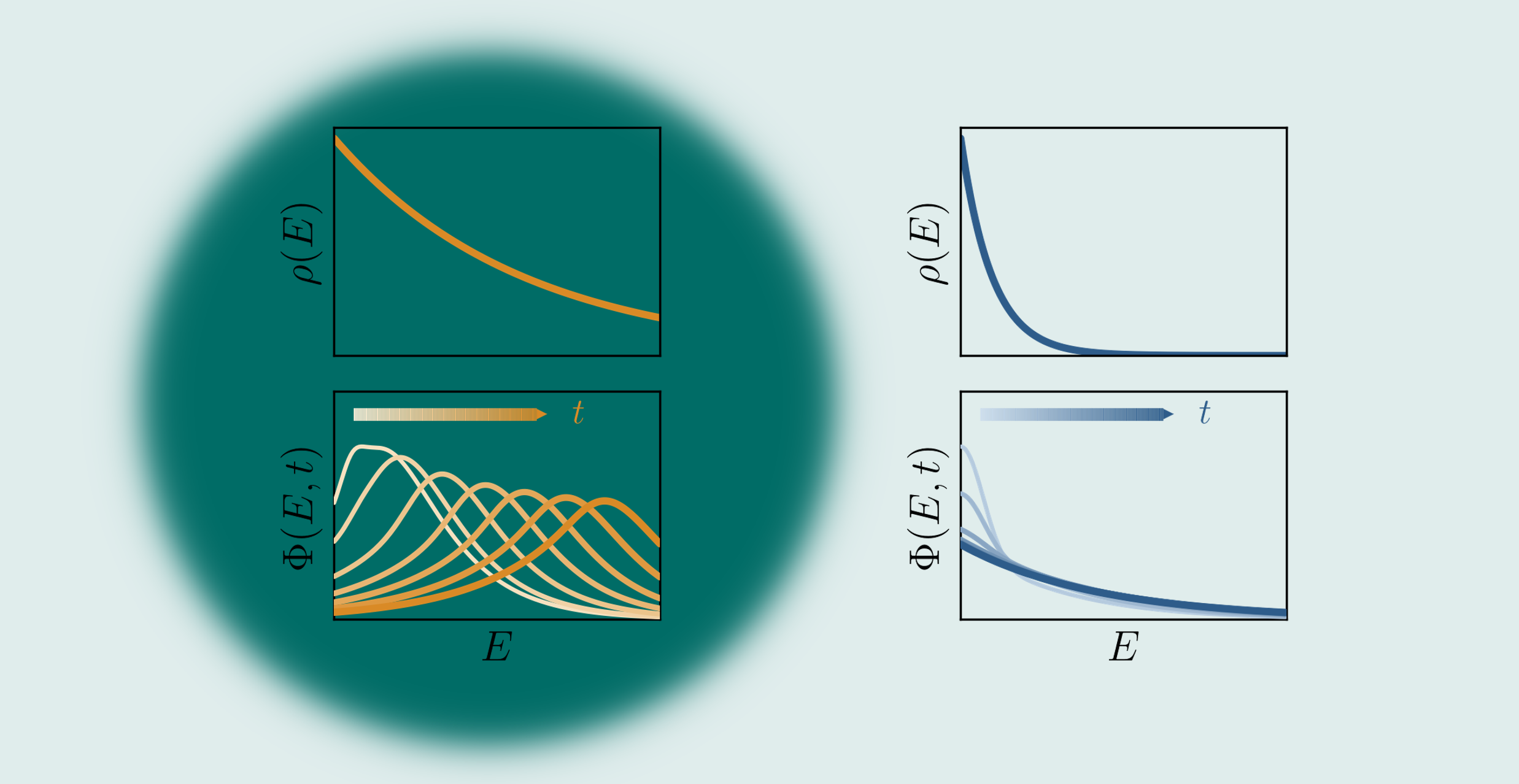}
    \caption{\textbf{Aging depends on local composition.}
    The segregation of solutes into high-density regions (green droplet) affects their internal aging dynamics, described by a trap model.
    Denser regions exhibit deeper traps, which implies aging dynamics with continuously increasing trap energies $E$ (left).
    In contrast, dilute regions have shallow traps and evolve toward an equilibrium distribution (right).
    }
    \label{M-fig:schematic}
\end{figure}

There are many models of phase separation with complex rheology~\cite{tanaka2022Viscoelastica,yoshimoto2021Viscoelastic,tree2017Multifluid,zhang2024Chemically}, but they typically do not exhibit aging dynamics.
Conversely, generic models of aging~\cite{Sollich1997,Lin2022,takaki2023Theory}, usually based on the seminal trap model~\cite{bouchaud1992Weak,monthus1996Models,bouchaud2000Aging}, rarely account for the dynamics of phase separation.
While trap models have been extended to include some spatial structure~\cite{bertin2003Subdiffusion,tapias2020Entropic}, no general theory yet combines a trap model with phase separation.

Here, we combine a trap model~\cite{bouchaud1992Weak,monthus1996Models,bouchaud2000Aging}, describing aging, with the seminal Flory--Huggins model~\cite{Flory1942,Huggins1941}, describing phase separation.
We show that aging affects phase coexistence even before glassy states become accessible.
Moreover, we uncover qualitative effects on nucleation, which can proceed via two steps, in which the glass is reached through an intermediate liquid nucleus.

\mysection{Results}

\mysubsection{Model definition}%
We consider an isothermal, incompressible mixture of solutes that can segregate from some solvent.
To account for aging, we assume that solute molecules can be in various states distinguished by a binding energy~$E\ge 0$.
The system's state is then characterized by the volume fraction $\Phi(E, \boldsymbol{x})$ of solutes in  state~$E$ at position $\boldsymbol{x}$.
Consequently, the total solute fraction is $\phi(\boldsymbol{x})=\int_0^\infty\Phi(E,\boldsymbol{x})\dd E$, whereas the solvent fraction is $\phiS(\boldsymbol{x})=1-\phi(\boldsymbol{x})$.
The thermodynamics of the system are described by the free energy density
\begin{align}
    f[\Phi] &= l\,\phiS\log\phiS + \chi\,\phi\,\phiS
    \notag\\
    &\quad + \int_0^\infty\!\dd E\,\Phi(E)\bigl[\log\Phi(E) + (\alpha(\phi) - \beta)\,E\bigr]
    \label{M-eqn:f_full}
\end{align}
in units of the thermal energy $k_\mathrm{B}T$ per molecular size of the solutes.
Here, the first term accounts for the translational entropy of the solvent, which increases proportional to the relative molecular size~$l$ of the solute with respect to the solvent.
The second term captures interactions between solvent and solutes, which are quantified by a Flory parameter~$\chi$~\cite{Flory1942,Huggins1941}.
Finally, the integral accounts for the internal states of the solutes, and we will show below that it leads to aging when the crucial parameter $\alpha$ is smaller than the inverse temperature~$\beta=1/k_\mathrm{B}T$.
While this model can be derived from a microscopic sticker picture (Appendix), we here study it as a simple model combining phase separation and aging.

The free energy~$f$ given by \Eqref{M-eqn:f_full} essentially combines a Flory--Huggins free energy with a density-dependent trap model.
To see this, we first consider the limit $\alpha\to\infty$, so that all solutes occupy state $E=0$ and $\Phi(E)\to\phi\delta(E)$.
In this first limit, $f$ reduces to Flory--Huggins form, $f_\mathrm{FH}(\phi) = \phi\log\phi + l\,\phiS\log\phiS + \chi\,\phi\,\phiS$, implying that the solutes phase separate from the solvent for sufficiently large interaction $\chi$~\cite{Flory1942,Huggins1941}.
To reveal the trap model, we consider a constant solute fraction~$\phi$, so the first two terms in \Eqref{M-eqn:f_full} become irrelevant constants.
The system's behavior is then governed by the last term, which corresponds to the effective free energy of the standard trap model with density of states $\rho(E)\propto e^{-\alpha(\phi) E}$ (Appendix).
In particular, $f$ is a Lyapunov functional of the passive dynamics obeying detailed balance~\cite{bouchaud1992Weak,monthus1996Models}.
Consequently, the system converges toward the equilibrium distribution 
$\Phi_\mathrm{eq}(E)\propto\rho(E)\,e^{\beta E}$
if the inverse temperature is smaller than the inverse glass temperature, $\beta < \alpha(\phi)$.
In contrast, $\Phi_\mathrm{eq}$ is not normalizable when $\beta > \alpha(\phi)$.
Instead, the energy density~$f$ decreases indefinitely, indicating aging~\cite{bouchaud2000Aging}.
Taken together, this suggest that the general case of finite $\alpha$ and spatially dependent $\phi$ describes a system that combines phase separation with aging modeled by a trap model.

\mysubsection{Long-term behavior}%
To reveal the behavior of the free energy given by \Eqref{M-eqn:f_full}, we first analyze its minima.
Since the system exhibits aging for $\alpha(\phi) < \beta$, such minima only exist in a restricted parameter regime.
For simplicity, we focus on systems where $\alpha(\phi)$ is a monotonically decreasing function of $\phi$, which describes solutes where deeply bound states are more abundant in regions of higher density~$\phi$.
In this case, the inverse glass temperature~$\alpha$ can be converted to a glass-transition volume fraction $\phig = \alpha^{-1}(\beta)$, such that the system ages toward a glassy state when $\phi>\phig$.
To minimize $f$ for liquid states ($\phi<\phig$), we substitute $\Phi_\mathrm{eq}(E)$ into $f$ to obtain
\begin{align}
    f_1(\phi) = f_\mathrm{FH}(\phi) + \phi\log\!\bigl[\alpha(\phi)-\beta\bigr]
    \,,
    \label{M-eqn:f_reduced}
\end{align}
which describes the system assuming internal states are equilibrated.
This equation essentially combines the standard Flory--Huggins free energy~$f_\mathrm{FH}$ with a destabilizing contribution that diverges at the glass transition $\phi=\phig$. %
In particular, $f_1$ describes a binary mixture with density~$\phi$ in the liquid-like regime ($\phi<\phig$).

\begin{figure*}[t]
    \centering
    \includegraphics[width=\textwidth]{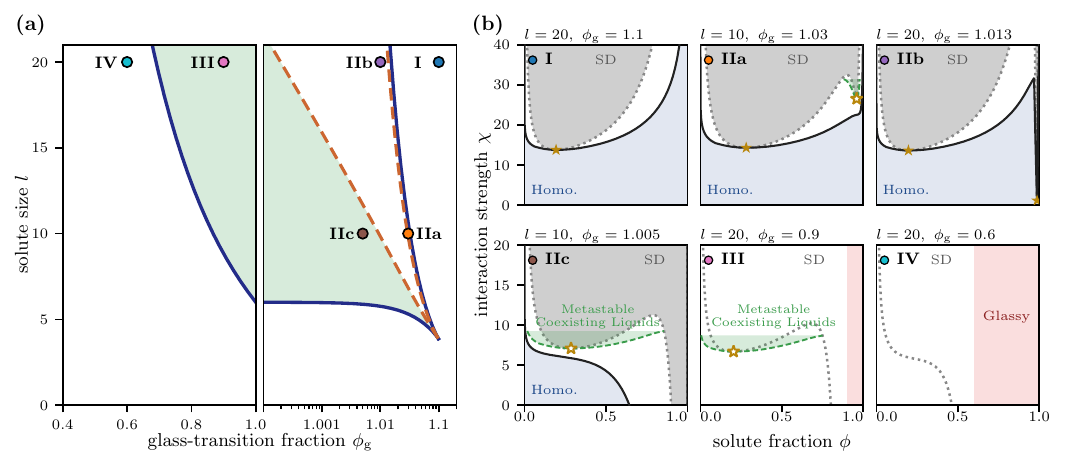}
    \caption{\textbf{Aging reshapes phase separation.}
    (a)~Meta phase diagram as a function of the glass-transition fraction $\phig$ and solute size $l$.
    The spinodal of the homogeneous state, $\chi(\phi)$, has multiple stationary points, indicating complex behavior, inside the solid blue boundary.
    The dashed brown lines mark parameters where critical points change stability.
    These lines divide the plane into six categories (I, IIa, IIb, IIc, III and IV), detailed in panel b.
    (b)~Phase diagrams as a function of solute fraction $\phi$ and interaction strength $\chi$ for exemplary parameters $(\phig, l)$ for the six categories indicated in panel~a.
    The red regions ($\phi \ge \phig$) mark the onset of aging without an equilibrium.
    For $\phi<\phig$, the phase behavior follows from \Eqref{M-eqn:f_reduced}, where equilibrium states are homogeneous below the binodal (solid line).
    Homogeneous mixtures are unstable above the spinodal (dotted line) and undergo spinodal decomposition (SD, gray region). %
    Critical points are marked by filled star if stable, and open stars if metastable.
    The green region indicates metastable coexisting liquids.
    }
    \label{M-fig:phase_diagrams}
\end{figure*}

For concreteness, we consider the lowest-order expansion of the inverse glass temperature, $\alpha(\phi) = \beta_0 + \theta(1-\phi)$, which can also be derived from the microscopic sticker model (Appendix).
Here, $\beta_0$ marks the inverse glass transition temperature for a pure solute system, whereas $\theta$ quantifies the increase of $\alpha$ when solvent is added.
Consequently, the glass-transition fraction becomes $\phig = 1-(\beta-\beta_0)/\theta$, reachable only when $\phig\in[0,1]$, i.e., when $0<\beta-\beta_0<\theta$, and the reduced energy given by \Eqref{M-eqn:f_reduced} becomes $f_1(\phi) = f_\mathrm{FH}(\phi) + \phi\log[\theta(\phig-\phi)]$.
Using this linear model, we next analyze the system as a function of density~$\phi$ and the model parameters $\phig$, $\chi$, and $l$.

We categorize the effects of glassiness by first studying the stability of homogeneous states.
Such states are stable below a critical interaction strength $\chisp(\phi)$, which we obtain from the spinodal condition $f_1''(\phi) = 0$.
Since the extrema of $\chisp(\phi)$ are the roots of $f_1'''$, which depends on $\phig$ and $l$, but not $\chi$, we use the number of monotonic branches of $\chisp$ as a signature to categorize the phase diagrams (\Figref{M-fig:phase_diagrams}a):
for $\phig>1$ the spinodal has two branches and one critical point (cat.~I) or four branches and two critical points (cat.~II).
For $\phig\le 1$ the glassy region enters the physical domain.
We then have three branches with one critical point (cat.~III), or only one branch if $\phig$ becomes sufficiently small (cat.~IV).
We further classify the categories by the stability of the critical points.
By inspecting the tangent line at the critical points, we find that the critical point in cat.~I is always stable and the one of cat.~III is always metastable.
The two critical points of cat.~II can be either both stable (cat.~IIb), or one stable and one metastable (cat.~IIa and cat.~IIc), leading to three subcategories (dashed boundaries in \Figref{M-fig:phase_diagrams}a).
The meta phase diagram shown in \Figref{M-fig:phase_diagrams}a already suggests that glassiness can impact the behavior substantially.

The influence of glassiness on equilibrium behavior becomes clear from the phase diagrams of the various categories, which we obtain via the common-tangent construction on $f_1$ (\Figref{M-fig:phase_diagrams}b).
When $\phig$ is large (cat.~I), the glass term $\phi\log[\theta(\phig-\phi)]$ is nearly linear in $\phi$.
Since a linear term does not change the convexity of $f_1$, the associated phase diagram is unaltered and exhibits dilute--dense coexistence for sufficiently large $\chi$.
Decreasing $\phig$ toward one, the glass term gains curvature and a second critical point appears on the dense side (cat.~II).
At the beginning (cat.~IIa), this dense critical point is metastable: two relatively dense liquids can coexist metastably, while the dilute--dense coexistence remains the global minimum (green shading in IIa).
Decreasing $\phig$ further (cat.~IIb) results in two separate binodals since both critical points are stable.
Further decreasing $\phig$ (cat.~IIc) changes the dilute critical point to be metastable.
Consequently, there is now a considerable window of metastable coexisting liquids in the $(\phi,\chi)$ phase diagram, while the global minimum is a coexistence between the dilute liquid and a very dense liquid at $\phi\approx 1$.
This picture persists for $\phig\le 1$, where we now find a region without equilibrium for $\phi\ge\phig$ since the glass term diverges as $\phi\to\phig$.
Since the free energy in this glassy region is arbitrarily negative, the phase diagram no longer has a well-defined binodal.
Instead, the system tends toward a coexistence between pure solvent and a glass, whose composition depends on the kinetic pathway.
However, the metastable liquid--liquid coexistence still survives in cat.~III (for sufficiently large $\phig\le 1$). 
For smaller $\phig$ (cat.~IV), the metastable state vanishes and the glass dominates the entire dense side.
Note that this sequence of phase behavior with decreasing $\phig$ requires a sufficiently large $l$, suggesting that metastable coexisting liquids require an entropic asymmetry between solvent and solute.
This is consistent with previous results~\cite{lutsko2006Theoretical}.
For $l<3.9$, $\phig$ only switches the system between a typical Flory--Huggins mixture (cat.~I) and one with a possible glassy state (cat.~IV).
In contrast, the phase diagrams suggest that larger $l$ enable interesting metastable states that could influence nucleation behavior.
In particular, the metastable liquid--liquid coexistence in cat.~III hints at two-step nucleation toward a glassy state, as previously observed for large molecules~\cite{Wolde1997,lutsko2006Theoretical,savage2009Experimental}.

\begin{figure*}[t]
    \centering
    \includegraphics[width=\textwidth]{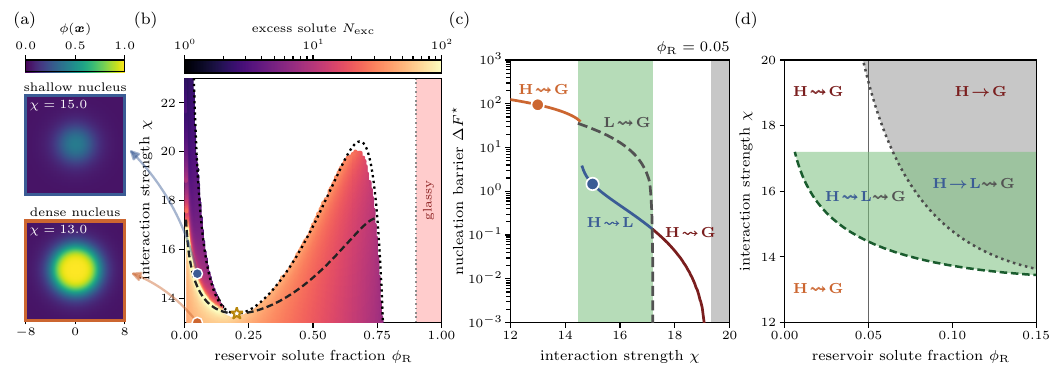}
    \caption{\textbf{Droplets can form and age via two-step nucleation.}
    (a) Density profiles of a shallow critical nucleus (top, blue frame) and a dense critical nucleus (bottom, orange frame) for reservoir solute fraction $\phiR = 0.05$.
    (b) Excess solute $N_\mathrm{exc}$ of critical nuclei as a function of $\phiR$ and interaction strength $\chi$, overlayed with the $(\chi,\phi)$ phase diagram in \Figref{M-fig:phase_diagrams}b III.
    Arrows mark the two examples shown in panel a.
    The discontinuity in $N_\mathrm{exc}$ coincides with the boundary of the metastable coexisting liquids (dashed line).
    The homogeneous state is unstable above the spinodal (dotted line).
    The critical point (star) is metastable.
    (c) Nucleation barrier $\Delta F^\star$ as a function of $\chi$ for $\phiR=0.05$.
    The dense nucleus (orange) and the shallow nucleus (blue) can both lead to a glass; within the metastable window (transparent green) the shallow nucleus first forms a liquid, requiring a secondary nucleation (gray dashed) whose barrier vanishes at the window's upper edge, beyond which the shallow nucleus reaches the glass directly (dark red).
    Disks mark examples in panel a.
    (d) Summary of nucleation behavior as a function of $\phiR$ and $\chi$.
    A homogeneous state (H) transitions to a glass (G), potentially via a liquid (L).
    Transitions happen either by instabilities ($\to$) or via nucleation ($\leadsto$).
    Lines mark the spinodal (dotted), the boundary of the metastable coexisting liquids (dashed) and $\phiR=0.05$ shown in panel c (vertical solid line).
    (a--d) Critical nuclei were determined by gentlest-ascent dynamics in two dimensions for $\phig=0.9$, $l=20$, $\theta=1$, $\kappa=4$ and $\vartheta=0.5$.
    }
    \label{M-fig:barriers}
\end{figure*}
\mysubsection{Nucleation}%
To study nucleation, we next determine the transition states, which are index-1 saddles of the free energy.
We show in the Appendix that these saddles can be obtained by only considering the first two moments, $\phi(\boldsymbol{x})$ and $\varepsilon(\boldsymbol{x})=\int_0^\infty\!\dd E\,E\,\Phi(E,\boldsymbol{x})$, of the full distribution $\Phi(E, \boldsymbol{x})$.
The associated reduced free energy density,
\begin{align}
f_2(\phi, \varepsilon) &= f_\mathrm{FH}(\phi) + \phi\log(\phi/\varepsilon) - \phi + [\alpha(\phi) - \beta]\varepsilon
    \;,
\end{align}
captures all stationary points of the full free energy, essentially because the excess free energy only depends on the two moments~\cite{Warren1998,Sollich1998,Sollich2001,Sollich2002}.
In contrast to $f_1$, which diverges at $\phig$, $f_2$ stays finite there because $\varepsilon$ is an independent variable, so it also describes compositions beyond $\phig$.
To enable smooth spatial profiles, we extend this free energy to spatially inhomogeneous systems by adding interfacial terms for both fields,
\begin{align}
    F_2 &= \int\!\dd V \,\Bigl\{ f_2(\phi,\varepsilon) + \frac{\kappa}{8\phi}|\bnabla\phi|^2 + \frac{\vartheta}{8\varepsilon}|\bnabla\varepsilon|^2 \Bigr\}
    \,,
    \label{M-eqn:F_full_func}
\end{align}
where the volume $V$ is measured in units of the solute's molecular size.
The subscript indicates that $F_2$ depends on $\Phi(E,\boldsymbol{x})$ only through its first two moments.
The free energy resolving the full distribution is given in the Appendix.
Here, $\kappa$ and $\vartheta$ penalize gradients in the respective fields $\phi$ and $\varepsilon$, which gives rise to surface tension in classical mixtures~\cite{Cahn1958}.
Weighting each squared gradient by the inverse density follows the interfacial free energy of copolymer melts~\cite{tang1991Free}, since both $\phi$ and $\varepsilon$ are proportional to monomer densities (Appendix).
For simplicity, we consider a grand-canonical ensemble, where the system is in contact with a  reservoir at chemical potential $\mu = f_1'(\phiR)$ and corresponding reservoir fraction $\phiR$, while $\varepsilon$ is an unconstrained internal variable (Appendix).
We determine the saddles of $F_2$ in two dimensions using gentlest-ascent dynamics by descending $F_2$ in every direction but the direction of smallest curvature~\cite{e2011Gentlest,yin2019HighIndexa,gander1989Constrained,zhou2019Linear}.
\Figref{M-fig:barriers}a shows examples of the saddles, which are the critical nuclei for the transition from homogeneous to phase separated states.

We next focus on nucleation in the metastable coexisting liquids in the cat.~III phase diagram (\Figref{M-fig:phase_diagrams}b) by numerically determined critical nuclei as a function of $\phiR$ and $\chi$.
\Figref{M-fig:barriers}b shows that such nuclei are only present in the nucleation-and-growth regime below the spinodal (dotted line).
In this region, the excess solute $N_\mathrm{exc}\equiv\int\!\dd V\,[\phi(\boldsymbol{x})-\phiR]$ (as well as the maximal density; see Appendix) exhibits a discontinuity, indicating qualitatively different critical nuclei.
Indeed, for large $\phiR$ (or low interaction parameters~$\chi$), we observe dense nuclei with peak fractions reaching and even exceeding the glass transition ($\phi_\mathrm{max}\gtrsim\phig$) accompanied by a large excess solute (lower panel in \Figref{M-fig:barriers}a).
Note that the nucleus is not yet in the glassy regime since the second gradient term in \Eqref{M-eqn:F_full_func} penalizes spatial variation of $\varepsilon$.
Essentially, this gradient term limits the mean trap depth $\varepsilon/\phi$, so that 
the dense nucleus tends to reach local compositions above the homogeneous threshold $\phig$ before it becomes glassy.
In contrast, critical nuclei have peak fractions well below $\phig$ for stronger interactions (and only if the reservoir density $\phiR$ is low enough), which we call shallow nuclei (upper panel in \Figref{M-fig:barriers}a).
A stability analysis (dashed line in \Figref{M-fig:barriers}b) indicates that the discontinuity coincides with the metastable binodal.

The two classes of nuclei suggest qualitatively different nucleation pathways.
To see this, we quantify the nucleation barriers~$\Delta F^\star \equiv F_2[\phi_\mathrm{crit},\varepsilon_\mathrm{crit}] - V f_1(\phiR) - \mu N_\mathrm{exc}$, which is the grand-potential excess of the critical nucleus over the reservoir.
We generally observe the largest $\Delta F^\star$ (implying lowest nucleation rates, which scale with $\exp(-\Delta F^\star)$~\cite{classical_nucleation_theory}) for weak interactions~$\chi$, where the dense nuclei dominate (orange line in \Figref{M-fig:barriers}c).
Increasing $\chi$ generally lowers $\Delta F^\star$, which drops even more sharply in the parameter range of metastable liquids (green band).
In this parameter regime, shallow nuclei emerge (blue line), with even lower $\Delta F^\star$.
Consequently, the shallow nucleation path is more likely than that of the dense nuclei, indicating that the nucleation behavior also exhibits a discontinuity.
In the metastable region, the dilute phase first nucleates a dense liquid, whose composition is set by the dense branch of the metastable binodal (dashed green line in \Figref{M-fig:phase_diagrams}b).
However, this liquid phase is itself metastable and will thus eventually become a glass via a secondary nucleation event (gray dashed line in \Figref{M-fig:barriers}c).
The barrier associated with this secondary nucleation vanishes at the upper edge of the metastable window, beyond which the shallow nucleus directly transitions to a glass (dark red line in \Figref{M-fig:barriers}c).
Here, the peak density stays close to the reservoir fraction, and the barrier is orders of magnitudes lower than that of the dense nucleus at low~$\chi$.
Taken together, this detailed analysis shows that a glass can be reached via either a dense or a shallow nucleus.

We summarize the nucleation behavior of our model in \Figref{M-fig:barriers}d, which indicates how a homogeneous state (H) reaches a glass (G), potentially via an intermediate liquid state (L).
Below the spinodal (dotted line), the homogeneous state is metastable, and reaches the glass either by direct nucleation via a dense nucleus ($\text{H}\leadsto\text{G}$, at low $\chi$ or low $\phi$) or via a shallow nucleus that first forms a liquid ($\text{H}\leadsto\text{L}\leadsto\text{G}$, green region).
Above the metastable window, we again find direct nucleation ($\text{H}\leadsto\text{G}$, upper-left region), and this region is continuously connected to the low-$\chi$ direct-nucleation branch: at $\phi$ below the metastable binodal's smallest $\phi$, the barrier for $\text{H}\leadsto\text{G}$ varies smoothly with $\chi$, whereas at larger $\phi$ the same branch is interrupted by the metastable window and $\Delta F^\star$ jumps discontinuously across it.
This discontinuity might be viewed as a critical point of the family of index-1 saddles.
Above the spinodal, the homogeneous state exhibits spinodal decomposition either directly to a glass ($\text{H}\to\text{G}$) or to a metastable liquid, which then nucleates the glass ($\text{H}\to\text{L}\leadsto\text{G}$).
Taken together, our theory indicates that the interplay of phase separation and aging yields various nucleation pathways, where in particular an intermediate liquid nucleus can accelerate the formation of a glass.

\mysubsection{Discussion}%
We showed that coupling Flory--Huggins theory with a trap model leads to complex phase diagrams and a two-step nucleation pathway.
This two-step nucleation is reminiscent of the classical Warren scenario, where phase separation is followed by slow inter-diffusion~\cite{Warren1999}.
The associated separation of nucleation barriers may be biologically significant: efficient formation of functional liquid condensates requires a small barrier, while a large barrier delays the transition to pathological solid-like states associated with neurodegenerative disease~\cite{Alberti2021}.
This behavior is likely a generic consequence of combining phase separation and aging, but to make quantitative predictions, our model would need to be augmented with molecular-scale features, such as interfacial molecular orientation, which might trigger solidification~\cite{Shen2020,Emmanouilidis2024}.
Moreover, our model cannot characterize the final glassy state, which requires a full dynamical treatment of nucleation and growth; such an extension could also reveal how noise induces heterogeneity among droplets~\cite{erkamp2023Spatially}.
Natural extensions include incorporating hydrodynamics~\cite{Sollich1997}, as well as active processes that may hinder aging~\cite{makasewicz2025Formation,Zwicker2025}.
Beyond these theoretical extensions, our minimal framework may also help guide \textit{in vitro} efforts to study and engineer the hardening of condensates.

\textit{Acknowledgements---}We thank Peter Sollich and the entire Zwicker group for fruitful discussions.
We gratefully acknowledge funding from the Max Planck Society and the European Research Council (ERC, EmulSim, 101044662).

\textit{Data availability}---%
The data that support the findings of this study were generated by numerical simulations.
The source code of the simulations will be made publicly available on Zenodo.

\bibliography{refs}

\clearpage
\onecolumngrid
\appendix
\setcounter{secnumdepth}{2}

\setcounter{figure}{0}
\renewcommand{\thefigure}{A\arabic{figure}}
\renewcommand{\theHfigure}{A\arabic{figure}}
\renewcommand{\theHequation}{app.\Alph{section}.\arabic{equation}}

\bigskip
This Appendix describes the theoretical and numerical details of the model discussed in the main text.
In particular, \Secref{S-sec:si_micro} derives the model's free-energy density from a microscopic sticker picture.
\Secref{S-sec:si_reduced} then derives the two reduced free energy densities, $f_1(\phi)$ and $f_2(\phi,\varepsilon)$, used throughout the main text.
It shows that $f_1$ describes homogeneous equilibria exactly, and that the spatially extended free energy built from $f_2$ preserves the stationary points of the full model, including the critical nucleus.
\Secref{S-sec:si_metapd} gives the classification behind the meta phase diagram of Fig.~2a, and \Secref{S-sec:si_nucleus} derives the gentlest-ascent construction used to locate the critical nucleus shown in Fig.~3 in the main text.

\section{Microscopic derivation of the free energy}
\label{S-sec:si_micro}

The main text introduces the free-energy density
\begin{align}
    f[\Phi] = l\,\phiS\log\phiS + \chi\,\phi\,\phiS
      + \int_0^\infty\!\dd E\,\Phi(E)\bigl[\log\Phi(E) + (\alpha(\phi) - \beta)\,E\bigr]
    \label{S-eqn:f_full_maintext}
    \;,
\end{align}
which is identical to Eq.~(1) in the main text.
In contrast to the main text, where this equation is introduced phenomenologically, we here derive it from a microscopic model in four steps.
First, we cast the trap dynamics as a detailed-balanced chemical reaction network.
Next, we show that these dynamics admit a Lyapunov functional, which identifies the free energy associated with the trap dynamics.
Then, we use a microscopic sticker picture to explain the form $\rho(E;\phi)\propto e^{-\alpha(\phi)E}$ and the particular $\alpha(\phi)$ used for concreteness in the main text.
Finally, we combine the trap free energy with the Flory--Huggins free energy to couple the traps to phase separation, and show that the result is still a Lyapunov functional of the trap dynamics.

\subsection{Trap dynamics form a detailed-balanced reaction network}
\label{S-sec:si_trap_reaction}

Let $\Phi(E,t)$ be the volume-fraction density of solute at trap depth $E\ge0$, and let $\rho(E)>0$ count the microstates at that depth.
Thermally activated hopping between depths is governed by the master equation
\begin{align}
    \frac{\partial \Phi(E)}{\partial t}
    = \int_0^\infty \dd E'\,
      \bigl[ W(E'\rightarrow E)\,\rho(E)\,\Phi(E')
           - W(E\rightarrow E')\,\rho(E')\,\Phi(E) \bigr]
    \;,
    \label{S-eqn:trap_master}
\end{align}
where the jump rate $W(E\rightarrow E')$ obeys detailed balance,
\begin{align}
    \frac{W(E\rightarrow E')}{W(E'\rightarrow E)} = e^{\beta\,(E' - E)}
    \;,
    \label{S-eqn:W_ratio}
\end{align}
at inverse temperature $\beta=1/\kBT$.
Expressing $W$ through a symmetric attempt-rate kernel $r(E,E')=r(E',E)\ge0$ as
\begin{align}
    W(E\rightarrow E') = \frac{r(E,E')}{\rho(E)\,\rho(E')}\,e^{-\beta E}
    \;,
    \label{S-eqn:W_from_r}
\end{align}
automatically satisfies \Eqref{S-eqn:W_ratio} for any symmetric $r$.
Substituting into \Eqref{S-eqn:trap_master} and introducing the effective chemical potential of the trap model
\begin{align}
    \mu_\mathrm{T}(E) \equiv \log\Phi(E) - \log\rho(E) - \beta E
    \;,
    \label{S-eqn:mu_def}
\end{align}
converts the master equation to the chemical-reaction form
\begin{align}
    \frac{\partial \Phi(E)}{\partial t}
    = \int_0^\infty \dd E'\, r(E,E')\,\bigl[e^{\mu_\mathrm{T}(E')} - e^{\mu_\mathrm{T}(E)}\bigr]
    \;.
    \label{S-eqn:trap_reaction}
\end{align}
This equation describes a detailed-balanced reaction network over depth levels~$E$.

\subsection{Lyapunov functional of the trap dynamics}
\label{S-sec:si_lyapunov}

Equations such as \Eqref{S-eqn:trap_reaction} admit a Lyapunov functional~\cite{schnakenberg1976Network}.
To derive it for \Eqref{S-eqn:trap_reaction}, we first seek stationary states, which obey $\partial_t\Phi=0$.
For any~$r$, this implies $\mu_\mathrm{T}(E)=\mathrm{const}$, giving the steady state
\begin{align}
    \Phi_\mathrm{ss}(E) \propto \rho(E)\,e^{\beta E}
    \;.
    \label{S-eqn:Phi_ss}
\end{align}
The relative entropy of $\Phi$ with respect to the unnormalized stationary profile $\rho(E)\,e^{\beta E}$,
\begin{align}
    f_\mathrm{T}[\Phi]
    &\equiv \int_0^\infty \dd E\, \Phi(E)\,\log\!\frac{\Phi(E)}{\rho(E)\,e^{\beta E}}
\notag \\
    &= \int_0^\infty \dd E\, \Phi(E) \bigl[ \log\Phi(E) - \log\rho(E) - \beta E \bigr]
    \;,
    \label{S-eqn:F_trap}
\end{align}
decreases monotonically along the dynamics.
To see this, we differentiate $f_\mathrm{T}$ along \Eqref{S-eqn:trap_reaction},
\begin{align}
    \frac{\dd f_\mathrm{T}}{\dd t}
    &= \int_0^\infty \dd E\, \mu_\mathrm{T}(E)\,\frac{\partial \Phi(E)}{\partial t}
\notag \\
    & = \iint_0^\infty \dd E\,\dd E'\, r(E,E')\, \mu_\mathrm{T}(E)\,
       \bigl[ e^{\mu_\mathrm{T}(E')} - e^{\mu_\mathrm{T}(E)} \bigr]
\notag \\
    &= -\frac{1}{2} \iint_0^\infty \dd E\,\dd E'\, r(E,E')\,
      \bigl[\mu_\mathrm{T}(E) - \mu_\mathrm{T}(E')\bigr]\,
      \bigl[e^{\mu_\mathrm{T}(E)} - e^{\mu_\mathrm{T}(E')}\bigr]
    \;\le\; 0
    \;,
    \label{S-eqn:H_theorem}
\end{align}
where the first line uses $\delta f_\mathrm{T}/\delta\Phi(E)=\mu_\mathrm{T}(E)+1$, whose constant part integrates to zero by mass conservation, and the last line symmetrizes in $E\leftrightarrow E'$ using $r(E,E')=r(E',E)$.
The inequality follows from $(a-b)(e^a-e^b)\ge0$, with equality iff $\mu_\mathrm{T}$ is constant, i.e., $\Phi\propto\Phi_\mathrm{ss}$.
$f_\mathrm{T}$ is therefore a Lyapunov functional of the trap dynamics, which is stationary only at $\Phi_\mathrm{ss}$ and strictly decreases otherwise.
When $\Phi_\mathrm{ss}(E)$ is not normalizable on $[0,\infty)$, no steady state exists, which is the aging regime discussed in the main text.

\subsection{Stickers make the state density composition dependent}
\label{S-sec:si_sticker}

A typical trap model takes $\rho(E)=e^{-\tilde\beta_0 E}$ with $\tilde\beta_0>0$, where $\tilde\beta_0$ is usually assumed to be constant.
Here, we derive how $\rho(E)$ acquires a composition dependence, which later couples the traps to phase separation.

We consider molecules carrying transient sticky sites, each in a bound or unbound state with binding enthalpy $-\epsilon_\mathrm{b}$ ($\epsilon_\mathrm{b}>0$).
We assume that the state density of the sticky-site count in the unbound reference state is a pure exponential, and that the depth is set by the number of bound sites, $E=n\epsilon_\mathrm{b}$.
Each sticky site can additionally form a transient crosslink with a partner site on another molecule, whose local density tracks the solute volume fraction.
We count binding configurations at fixed numbers $P_\mathrm{s}$ of sticky sites and $P_\mathrm{p}$ of partner sites, neglecting saturation, so that every partner site remains available.
With sticker and partner positions $\boldx_{\mathrm{s},i}$, $\boldx_{\mathrm{p},j}$ and density operators $\hat\rho_\mathrm{s}(\boldx)=\sum_i\delta(\boldx_{\mathrm{s},i}-\boldx)$, $\hat\rho_\mathrm{p}(\boldx)=\sum_j\delta(\boldx_{\mathrm{p},j}-\boldx)$, a crosslink requires sticker-partner contact within a volume $b^3$ and carries fugacity $z$.
Summing over the number of crosslinks $n_\mathrm{c}$ and over all pairings gives
\begin{align}
    N_\mathrm{binding}
    = \sum_{n_\mathrm{c}=0}^{\infty} \frac{z^{n_\mathrm{c}}}{n_\mathrm{c}!}\,
       \prod_{c=1}^{n_\mathrm{c}} \left[\sum_{i_c,j_c}\, b^3\, \delta\!\left(\boldx_{\mathrm{s},i_c} - \boldx_{\mathrm{p},j_c}\right)\right]
    = \exp\!\left[ z\, b^3 \int \dd\boldx\, \hat\rho_\mathrm{s}(\boldx)\,\hat\rho_\mathrm{p}(\boldx) \right]
    \;,
    \label{S-eqn:N_binding_exp}
\end{align}
using $\delta(\boldx_\mathrm{s}-\boldx_\mathrm{p})=\int\dd\boldx\,\delta(\boldx_\mathrm{s}-\boldx)\delta(\boldx_\mathrm{p}-\boldx)$ to write the pairwise sum as a product of densities, after which the series in $n_\mathrm{c}$ resums to an exponential.
For a molecule at depth $E=n\epsilon_\mathrm{b}$, the sum runs over its $n$ bound sticky sites, and at mean-field level the partner density is proportional to $\phi$.
The number of binding configurations then becomes
\begin{align}
    N_\mathrm{binding}(E) \propto e^{\theta E\phi} = e^{\theta E(1-\phiS)}
    \;,
    \label{S-eqn:N_binding_mf}
\end{align}
with the mean-field sticker-partner coupling $\theta\propto z\,b^3/\epsilon_\mathrm{b}$.
Combining this with the bare exponential state density gives the composition-dependent state density
\begin{align}
    \rho(E;\phi) \propto e^{-(\tilde\beta_0-\theta\phi)E} = e^{-\alpha(\phi)E}
    \;,\qquad
    \alpha(\phi) = \beta_0 + \theta(1-\phi)
    \;,
    \label{S-eqn:rho_micro}
\end{align}
with $\beta_0\equiv\tilde\beta_0-\theta$.
The $E$-independent prefactor of $\rho$ does not depend on $\phi$, so it only adds a term linear in $\phi$ to the free energy, which does not affect the phase behavior.
This is one way to obtain the linear form of $\alpha(\phi)$ that we use for concreteness in the main text.
Physically, solvent shifts the trap landscape toward shallower traps, since a larger $\phiS$ leaves fewer sticky partners available and thus suppresses deeply bound configurations.

\subsection{Flory--Huggins terms couple the traps to phase separation}
\label{S-sec:si_coupling}

Physically, \Eqref{S-eqn:F_trap} comprises the entropy of the trap distribution and the energy of occupying traps.
If the solute had only a single trap state, $f_\mathrm{T}$ would reduce to the ideal-gas entropy $\phi\log\phi$ of the solute, up to a term linear in $\phi$ that does not affect the phase behavior.
We therefore couple the traps to phase separation by adding the translational entropy of the solvent, $l\,\phiS\log\phiS$, and the Flory--Huggins interaction, $\chi\,\phiS\,\phi$, which gives the homogeneous free-energy density (per solute molecular volume $\nu$, in units of $\kBT$)
\begin{align}
    f[\Phi] = l\,\phiS\log\phiS + f_\mathrm{T}[\Phi] + \chi\,\phiS\,\phi
    \;,
    \label{S-eqn:f_fht}
\end{align}
where $l=\nu/\nu_\mathrm{S}$ is the relative molecular size and $\phiS=1-\phi$ due to the no-void constraint.
With $-\log\rho(E;\phi)=\alpha(\phi)E$ from \Eqref{S-eqn:rho_micro}, \Eqref{S-eqn:f_fht} reproduces \Eqref{S-eqn:f_full_maintext} exactly.
The trap dynamics of \Eqref{S-eqn:trap_reaction} conserve $\phi=\int\Phi\,\dd E$, and the added Flory--Huggins terms depend on $\Phi$ only through $\phi$, so they stay constant along the dynamics.
Hence $f$ remains a Lyapunov functional of the trap dynamics, with steady state
\begin{align}
    \Phi_\mathrm{eq}(E) \propto \rho(E;\phi)\,e^{\beta E} = e^{-[\alpha(\phi)-\beta]E}
    \;.
    \label{S-eqn:Phi_eq_fht}
\end{align}
This distribution is normalizable on $[0,\infty)$ iff $\alpha(\phi)>\beta$.
For the monotonically decreasing $\alpha(\phi)$ used in the main text, this condition defines the glass-transition fraction $\phig=\alpha^{-1}(\beta)$, and the system ages for $\phi\ge\phig$.

\section{Derivation of reduced free energies}
\label{S-sec:si_reduced}

Analyzing the free energy density $f[\Phi]$ of \Eqref{S-eqn:f_full_maintext} is cumbersome, since it depends on the entire depth distribution $\Phi(E)$.
Instead, we fix a few low-order moments of $\Phi(E)$ and minimize $f$ exactly over all distributions that share them.
We here derive the two forms used in the main text.
The first, $f_1(\phi)$, retains only the solute fraction $\phi$.
This suffices in homogeneous systems, where the trap distribution can freely adopt its optimal shape at the given composition.
We use $f_1$ for phase diagrams and spinodals (\Secref{S-sec:si_metapd}).
The second, $f_2(\phi,\varepsilon)$, additionally retains the first moment $\varepsilon$.
Since $\varepsilon$ enters the interfacial free energy of \Secref{S-sec:si_interface}, this form is convenient for spatially varying systems.
We use $f_2$ to locate critical nuclei (\Secref{S-sec:si_nucleus}).

\subsection{One-field form \texorpdfstring{$f_1(\phi)$}{f₁(φ)} for homogeneous states}
\label{S-sec:si_f1}

To reduce \Eqref{S-eqn:f_full_maintext} to a function of the solute fraction alone, we minimize it over $\Phi(E)\ge0$ while holding $\phi=\int\Phi\,\dd E$ fixed.
We impose this constraint with a Lagrange multiplier $\lambda_0$ and demand that the variation of $f-\lambda_0\int\Phi\,\dd E$ with respect to $\Phi(E)$ vanishes,
\begin{align}
    \log\Phi(E) + 1 + [\alpha(\phi)-\beta]\,E - \lambda_0 = 0
    \;.
    \label{S-eqn:f1_stationarity}
\end{align}
Solving for $\Phi(E)$ and fixing $\lambda_0$ by the constraint yields the normalized exponential
\begin{align}
    \Phi^\star(E) = \phi\,[\alpha(\phi)-\beta]\,e^{-[\alpha(\phi)-\beta]E}
    \;,
    \label{S-eqn:Phi_star}
\end{align}
which exists only for $\alpha(\phi)>\beta$, i.e., $\phi<\phig$.
Beyond the glass line, for $\phi>\phig$, the exponential is not normalizable and the system ages (\Secref{S-sec:si_lyapunov}).
Inserting \Eqref{S-eqn:Phi_star} back into \Eqref{S-eqn:f_full_maintext} gives
\begin{align}
    f_1(\phi) = f_\mathrm{FH}(\phi) + \phi\log[\alpha(\phi)-\beta]
    \;,\qquad
    f_\mathrm{FH}(\phi) = \phi\log\phi + l\,\phiS\log\phiS + \chi\,\phi\,\phiS
    \;,
    \label{S-eqn:f_reduced}
\end{align}
matching the main text.

\mypar{$f_1$ captures homogeneous equilibria and their stability exactly}%
We next show that $f_1$ reproduces both the equilibria and the stability of homogeneous states of the full model without approximation, which allows us to study the homogeneous phase behavior using $f_1$ alone (\Secref{S-sec:si_metapd}).
Since \Eqref{S-eqn:Phi_star} is the unique minimizer of $f$ at each fixed $\phi$, $f_1(\phi)$ is the lowest free energy of all homogeneous states with composition $\phi$.
Wherever a global equilibrium exists, the common-tangent construction on $f_1$ therefore yields the coexisting phases of the full model exactly.
The same argument determines the stability.
A perturbation that leaves $\phi$ unchanged merely redistributes $\Phi$ at fixed composition and therefore raises $f$.
The only direction that can become unstable thus changes $\phi$, and its curvature is $f_1''(\phi)$.
Consequently, $f_1''(\phi)=0$ locates the spinodal of $f$ exactly.

\subsection{Two-field form \texorpdfstring{$f_2(\phi,\varepsilon)$}{f₂(φ,ε)} for spatially varying states}
\label{S-sec:si_f2}

We next additionally keep the first moment of the trap distribution,
\begin{align}
    \varepsilon \equiv \int_0^\infty \dd E\, E\,\Phi(E)
    \;,
    \label{S-eqn:eps_first_moment}
\end{align}
fixed, which we can also express through the mean trap depth $\mathcal{E}\equiv\varepsilon/\phi$.
The two constraints require two Lagrange multipliers, $\lambda_0$ and $\lambda_1$, and the variation of $f-\lambda_0\int\Phi\,\dd E-\lambda_1\int E\,\Phi\,\dd E$ with respect to $\Phi(E)$ vanishes when
\begin{align}
    \log\Phi(E) + 1 + [\alpha(\phi)-\beta]\,E - \lambda_0 - \lambda_1 E = 0
    \;.
    \label{S-eqn:f2_stationarity}
\end{align}
Since $\phi$ is held fixed, the term $[\alpha(\phi)-\beta]E$ is itself linear in $E$ and can be absorbed into $\lambda_1$.
The solution is therefore again exponential, with amplitude and decay rate fixed by the two constraints,
\begin{align}
    \Phi^\star(E) = \frac{\phi}{\mathcal{E}}\,e^{-E/\mathcal{E}}
    \;,\qquad
    \mathcal{E}=\frac{\varepsilon}{\phi}
    \;,
    \label{S-eqn:Phi_star_2field}
\end{align}
where the decay rate $1/\mathcal{E}$ is set by the imposed mean depth rather than by $\alpha(\phi)-\beta$.
\Eqref{S-eqn:Phi_star_2field} is therefore normalizable for any finite $\varepsilon>0$, in particular also for $\phi>\phig$, where \Eqref{S-eqn:Phi_star} ceases to exist.
Inserting \Eqref{S-eqn:Phi_star_2field} back into \Eqref{S-eqn:f_full_maintext} gives
\begin{align}
    f_2(\phi,\varepsilon) = f_\mathrm{FH}(\phi) + \phi\log(\phi/\varepsilon) - \phi + [\alpha(\phi)-\beta]\,\varepsilon
    \;,
    \label{S-eqn:f_2field}
\end{align}
matching the main text.
Minimizing $f_2$ over $\varepsilon$ returns $\varepsilon^\star(\phi)=\phi/[\alpha(\phi)-\beta]$, the first moment of \Eqref{S-eqn:Phi_star}, and recovers $f_1(\phi)$, so that $\min_\varepsilon f_2=f_1$ for $\phi<\phig$.

\mypar{$f_2$ locates critical nuclei exactly}%
We next show that the critical nuclei of the full model can be determined from $f_2$ without approximation.
To this end, we write the spatially extended free energy, resolved in $\Phi(E,\boldx)$, as
\begin{align}
    F[\Phi]
    = \int \dd\boldx \left\{
        l\,\phiS\log\phiS + \chi\,\phiS\,\phi
        + \int_0^\infty\!\dd E\,\Phi(E,\boldx)\bigl[\log\Phi(E,\boldx) + \bigl(\alpha(\phi) - \beta\bigr)E\bigr]
      \right\}
      + F_\mathrm{int}[\phi,\varepsilon]
    \;,
    \label{S-eqn:F_full_level}
\end{align}
where $\phi(\boldx)$ and $\varepsilon(\boldx)$ are the local moments of $\Phi(E,\boldx)$, and $F_\mathrm{int}[\phi,\varepsilon]$ is the interfacial free energy given in \Eqref{S-eqn:F_int}.
Since $F_\mathrm{int}$ depends on $\Phi$ only through $\phi$ and $\varepsilon$, it is unaffected by changes of the shape of $\Phi(E,\boldx)$ at fixed moment fields.
With respect to these shape changes, $F$ is strictly convex.
To see this, consider a perturbation $\delta\Phi(E,\boldx)$ that leaves both moments unchanged, $\int\delta\Phi\,\dd E=\int E\,\delta\Phi\,\dd E=0$.
Only the entropy term $\int\Phi\log\Phi\,\dd E$ changes, since all other terms depend on $\Phi$ only through $\phi$ and $\varepsilon$, and its second variation,
\begin{align}
    \delta^2 F = \int\dd\boldx\int_0^\infty\dd E\,\frac{[\delta\Phi(E,\boldx)]^2}{\Phi(E,\boldx)}
    \;,
    \label{S-eqn:shape_convexity}
\end{align}
is positive for any nonzero $\delta\Phi$.
Among all distributions with prescribed moment fields $\phi(\boldx)$ and $\varepsilon(\boldx)$, $F$ is therefore minimized by the unique distribution $\Phi^\star$ of \Eqref{S-eqn:Phi_star_2field}, applied at each position.
Inserting this minimizer into \Eqref{S-eqn:F_full_level} gives the reduced free energy
\begin{align}
    F_2[\phi,\varepsilon] = \int\dd\boldx\, f_2\bigl(\phi(\boldx),\varepsilon(\boldx)\bigr) + F_\mathrm{int}[\phi,\varepsilon]
    \;,
    \label{S-eqn:F2_reduced}
\end{align}
which is the free energy of \Eqref{S-eqn:F_full_func} and Eq.~(4) in the main text.
Eliminating a variable over which the free energy is strictly convex has two consequences.
First, the stationary points of $F_2$ and $F$ correspond one to one, since every stationary point of $F$ is in particular stationary with respect to the shape and hence has $\Phi=\Phi^\star$.
Second, the eliminated shape directions are all stable, so the Hessian of $F_2$, which is the Schur complement of the Hessian of $F$ with respect to these directions, has the same number of unstable directions as the Hessian of $F$.
The same argument applies to the grand potentials $F-\mu\int\dd\boldx\,\phi$ and $\bar F_2=F_2-\mu\int\dd\boldx\,\phi$ used in \Secref{S-sec:si_grandcanonical}, since the added term also depends on $\Phi$ only through $\phi$.
Consequently, the index-one saddles of $\bar F_2$, which we determine in \Secref{S-sec:si_nucleus}, are exactly the critical nuclei of the full model.

\section{Classification of the meta phase diagram}
\label{S-sec:si_metapd}

The meta phase diagram in Fig.~2a of the main text classifies the homogeneous phase behavior by two properties.
The first is the number of critical points, which, together with whether the glass transition lies in the physical range ($\phig\le1$), defines the main categories I--IV.
The second is the stability of each critical point, which only matters when there are two critical points and therefore splits category II into the subcategories IIa, IIb, and IIc.
Both properties follow from the spinodal $\chisp(\phi)$, defined by $f_1''(\phi;\chisp)=0$, which gives
\begin{align}
    \chisp(\phi) = \frac{1}{2}\left[\frac{1}{\phi}+\frac{l}{\phiS}-\frac{2\phig-\phi}{(\phig-\phi)^2}\right]
    \;.
    \label{S-eqn:chi_spinodal_aged}
\end{align}
The last term captures the destabilizing effect of the glass term near $\phig$.
Each local minimum of $\chisp(\phi)$ marks a critical point of the phase diagram.

\subsection{The number of critical points depends only on \texorpdfstring{$\phig$}{φg} and \texorpdfstring{$l$}{l}}
\label{S-sec:si_metapd_number}

Differentiating \Eqref{S-eqn:chi_spinodal_aged} with respect to $\phi$ gives $\dd\chisp/\dd\phi=\tfrac{1}{2}f_1'''(\phi;\phig,l)$ with
\begin{align}
    f_1'''(\phi;\phig,l) = -\frac{1}{\phi^2}+\frac{l}{\phiS^2}-\frac{3\phig-\phi}{(\phig-\phi)^3}
    \;.
    \label{S-eqn:f1_third}
\end{align}
The extrema of $\chisp$ are thus the roots of $f_1'''$, which divide $\chisp$ into monotonic branches.
Since $f_1'''$ does not depend on $\chi$, the number of branches, and hence the number of critical points, depends only on $\phig$ and $l$.
For $\phig>1$, the spinodal has either two branches and one critical point (category I) or four branches and two critical points (category II).
For $\phig\le1$, the glassy region enters the physical domain, and the spinodal has either one branch and no critical point (category IV) or three branches and one critical point (category III).

The solid boundary in Fig.~2a marks where an additional critical point appears.
This happens when a new minimum and maximum of $\chisp$ emerge together, i.e., when two adjacent roots of $f_1'''$ merge into a double root, where
\begin{align}
    f_1'''(\phi;\phig,l) = 0
    \qquad\text{and}\qquad
    f_1''''(\phi;\phig,l) = \frac{2}{\phi^3}+\frac{2l}{\phiS^3}-\frac{2(4\phig-\phi)}{(\phig-\phi)^4} = 0
    \;.
    \label{S-eqn:f1_fourth}
\end{align}
These two equations for the three variables $\phi$, $\phig$, and $l$ define a curve, whose projection onto the $(\phig,l)$ plane is the solid boundary in Fig.~2a.
Along this boundary, $l$ has a minimum at $l\approx3.87$ and $\phig\approx1.10$.
Physically, solutes smaller than this never exhibit an additional critical point, so varying $\phig$ only switches the system between categories I and IV.

\subsection{Stability of critical points defines the subcategories}
\label{S-sec:si_metapd_stability}

\begin{figure}
    \centering
    \includegraphics[width=0.72\linewidth]{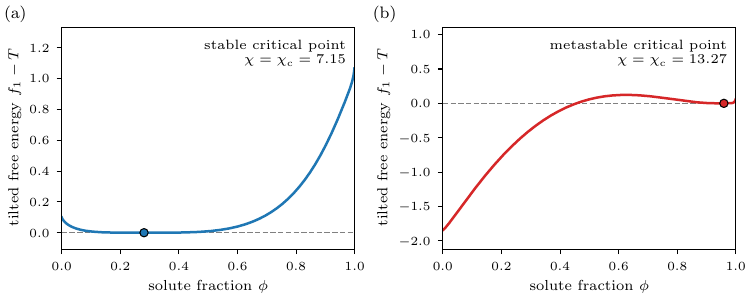}
    \caption{\textbf{A critical point is metastable when phase separation lowers the free energy.}
    Free energy $f_1$ at the critical interaction strength $\chi_\mathrm{c}$, minus its tangent $T$ at the critical composition, for the two critical points (disks) of category~IIa ($l=10$, $\phig=1.03$).
    (a)~The dilute critical point is stable, since $f_1-T\ge0$ everywhere.
    (b)~The dense critical point is metastable, since $f_1-T$ becomes negative at small $\phi$.}
    \label{S-fig:si_stability}
\end{figure}

A critical point at $(\phi_\mathrm{c},\chi_\mathrm{c})$ is stable if, at $\chi=\chi_\mathrm{c}$, the homogeneous state at $\phi_\mathrm{c}$ cannot lower its free energy by separating into coexisting phases.
To check this, we subtract from $f_1$ its tangent $T(\phi)$ at $\phi_\mathrm{c}$, which does not change phase equilibria because $T$ is linear in $\phi$.
The critical point is stable if the resulting tilted free energy $f_1-T$ is non-negative for all $\phi$.
It is metastable if $f_1-T$ becomes negative somewhere, since the system then lowers its free energy by separating into phases away from $\phi_\mathrm{c}$.
\Figref{S-fig:si_stability} illustrates both cases for the two critical points of category~IIa.
The stability changes where $f_1-T$ touches zero at a second point, which we determine numerically and which gives the dashed boundaries in Fig.~2a.

The single critical point of category~I is always stable.
In contrast, the single critical point of category~III is always metastable, because $f_1$ diverges to $-\infty$ as $\phi\to\phig\le1$, so $f_1-T$ always becomes negative.
The two critical points of category~II can both be stable (category~IIb), or either the dense one (category~IIa) or the dilute one (category~IIc) can be metastable.
Together with the number of critical points, this yields the six categories shown in Fig.~2a of the main text.

\section{Locating critical nuclei}
\label{S-sec:si_nucleus}

The critical nuclei in Fig.~3 of the main text are index-one saddle points of a spatially extended free energy.
Here, we specify this free energy and the method we use to locate its saddle points.

\subsection{Interfacial free energy}
\label{S-sec:si_interface}

A nucleus contains an interface, whose cost is not captured by the local free energy densities $f$ or $f_2$.
As in Cahn--Hilliard theory~\cite{Cahn1958}, we account for it by gradient terms that penalize spatial variations of the fields.
To motivate their form, we view the solution as a random copolymer in solvent.
Since $E=n\epsilon_\mathrm{b}$, the first moment $\varepsilon=\int E\,\Phi\,\dd E$ is proportional to the local density of bound stickers, which plays the role of the density of modified monomers.
For such copolymer melts, the squared gradient of each density is weighted by its inverse~\cite{tang1991Free}, which gives
\begin{align}
    F_\mathrm{int}
    = \int\dd\boldx\left[\frac{\kappa}{8\phi}|\bnabla\phi|^2+\frac{\vartheta}{8\varepsilon}|\bnabla\varepsilon|^2\right]
    \;,
    \label{S-eqn:F_int}
\end{align}
where $\kappa$ and $\vartheta$ set the stiffness of $\phi$ and $\varepsilon$, respectively.
Adding $F_\mathrm{int}$ to the two-field free energy density of \Eqref{S-eqn:f_2field} gives
\begin{align}
    F_2[\phi,\varepsilon] = \int\dd\boldx\left\{f_2(\phi,\varepsilon)+\frac{\kappa}{8\phi}|\bnabla\phi|^2+\frac{\vartheta}{8\varepsilon}|\bnabla\varepsilon|^2\right\}
    \;,
    \label{S-eqn:F_full_func}
\end{align}
which is Eq.~(4) of the main text.
As in the main text, volume is measured in units of the molecular volume $\nu$ of the solute.

\subsection{Grand-canonical ensemble}
\label{S-sec:si_grandcanonical}

We are interested in nucleation barriers in an infinitely large system, where a finite nucleus leaves the composition $\phiR$ of its surroundings unchanged.
The surroundings then act as a reservoir that exchanges solute with the nucleus, so we work in the grand-canonical ensemble with the grand potential
\begin{align}
    \bar F_2[\phi,\varepsilon] = F_2[\phi,\varepsilon] - \mu\int\dd\boldx\,\phi(\boldx)
    \;.
    \label{S-eqn:grand_potential}
\end{align}
The chemical potential $\mu$ is fixed by requiring that the uniform reservoir state $(\phiR,\varepsilon_\mathrm{R})$ is a stationary point of $\bar F_2$, which demands
\begin{align}
    \frac{\partial f_2}{\partial\varepsilon}\bigg|_{\phiR,\varepsilon_\mathrm{R}} = 0
    \qquad\text{and}\qquad
    \frac{\partial f_2}{\partial\phi}\bigg|_{\phiR,\varepsilon_\mathrm{R}} = \mu
    \;.
    \label{S-eqn:reservoir_conditions}
\end{align}
The first condition gives $\varepsilon_\mathrm{R}=\phiR/[\alpha(\phiR)-\beta]$, the first moment of \Eqref{S-eqn:Phi_star}, at which $f_2(\phiR,\varepsilon_\mathrm{R})=f_1(\phiR)$.
The second condition then gives $\mu=f_1'(\phiR)$, as in the main text.
A finite nucleation barrier further requires the reservoir to be locally stable, i.e., $\phiR<\phig$ and $f_1''(\phiR)>0$.
Here, $\mu$ is the chemical potential of the solute, not to be confused with the trap-level potential $\mu_\mathrm{T}(E)$ of \Eqref{S-eqn:mu_def}.

The excess grand potential of a profile relative to the reservoir is
\begin{align}
    \Delta F = \int\dd\boldx\left\{f_2(\phi,\varepsilon)-f_1(\phiR)-\mu\,(\phi-\phiR)+\frac{\kappa}{8\phi}|\bnabla\phi|^2+\frac{\vartheta}{8\varepsilon}|\bnabla\varepsilon|^2\right\}
    \;,
    \label{S-eqn:F_excess}
\end{align}
which vanishes for the reservoir.
The term $\mu\,(\phi-\phiR)$ integrates to $\mu N_\mathrm{exc}$, so $\Delta F$ evaluated at the critical nucleus is the barrier $\Delta F^\star$ defined in the main text.

The dense liquid formed by a shallow nucleus is itself metastable, and a glass can nucleate inside it.
We compute the barrier of this secondary nucleation, shown by the gray dashed line in Fig.~3c of the main text, inside an already phase-separated dense liquid rather than at the chemical potential of the original dilute reservoir.
To this end, we treat the dense liquid as the reservoir, set $\phiR$ to the composition of the dense branch of the metastable binodal at the same $\chi$, and evaluate $\Delta F$ of the dense nucleus for this reservoir.

\subsection{Gentlest-ascent dynamics}
\label{S-sec:si_gad}

According to \Secref{S-sec:si_f2}, the critical nucleus is an index-one saddle point of $\bar F_2$.
We locate it by gentlest-ascent dynamics (GAD)~\cite{e2011Gentlest,yin2019HighIndexa,gander1989Constrained,zhou2019Linear}, which ascends $\bar F_2$ along the direction of smallest curvature and descends along all others.
We work in the variables $\boldsymbol u=(\varphi,\varrho)^\top$ with $\varphi\equiv\sqrt\phi$ and $\varrho\equiv\sqrt\varepsilon$, in which the integrand of \Eqref{S-eqn:F_int} becomes $\tfrac{\kappa}{2}|\bnabla\varphi|^2+\tfrac{\vartheta}{2}|\bnabla\varrho|^2$.
This form is free of the singular factors $1/\phi$ and $1/\varepsilon$, which makes the numerics more stable.
We denote the first variation of $\bar F_2$ with respect to $\boldsymbol u$ by $\boldsymbol J=\delta\bar F_2/\delta\boldsymbol u$ and the corresponding Hessian by $\boldsymbol H$.
Products of fields such as $\boldsymbol v\cdot\boldsymbol H\boldsymbol v$ below denote spatial inner products, $\boldsymbol a\cdot\boldsymbol b\equiv\int\dd\boldx\,\sum_i a_i(\boldx)\,b_i(\boldx)$, where $i$ labels the two components.
GAD evolves a direction $\boldsymbol v$ toward the eigenvector of $\boldsymbol H$ with the smallest eigenvalue,
\begin{align}
    \frac{\partial\boldsymbol v}{\partial t} = -\boldsymbol H\boldsymbol v + \frac{\boldsymbol v\cdot\boldsymbol H\boldsymbol v}{\boldsymbol v\cdot\boldsymbol v}\,\boldsymbol v
    \;,
    \label{S-eqn:rq_dyn}
\end{align}
while the state descends along $-\boldsymbol J$ with its component along $\boldsymbol v$ reversed,
\begin{align}
    \frac{\partial\boldsymbol u}{\partial t} = -\boldsymbol J + 2\,\frac{\boldsymbol J\cdot\boldsymbol v}{\boldsymbol v\cdot\boldsymbol v}\,\boldsymbol v
    \;.
    \label{S-eqn:gad_dyn}
\end{align}
At a steady state of \Eqsref{S-eqn:rq_dyn}--\eqref{S-eqn:gad_dyn}, $\boldsymbol J=0$ and $\boldsymbol v$ is an eigenvector of $\boldsymbol H$, which our implementation (\Secref{S-sec:si_numerics}) selects as the one with the smallest eigenvalue.
We verify that this eigenvalue is negative, and for the nuclei shown in Fig.~3a of the main text that it is the only negative one.

\subsection{Numerical implementation}
\label{S-sec:si_numerics}

We compute critical nuclei in two dimensions.
Since the system is isotropic, we restrict the search to radially symmetric nuclei, $\boldsymbol u=\boldsymbol u(r)$, and do not consider nonradial perturbations.
The Laplacian and spatial integrals then reduce to
\begin{align}
    \nabla^2\boldsymbol u = \partial_r^2\boldsymbol u + \frac{1}{r}\,\partial_r\boldsymbol u
    \;,\qquad
    \int\dd\boldx\,g = \int_0^\infty 2\pi r\,g(r)\,\dd r
    \;.
    \label{S-eqn:radial_ops}
\end{align}
To represent the unbounded half-line with a finite number of grid points, we map $r$ onto $\xi\in(-1,1)$ via
\begin{align}
    r(\xi) = \frac{L\,\xi}{1-\xi^2}
    \;,
    \label{S-eqn:radial_map}
\end{align}
so that $\xi\to\pm1$ corresponds to $r\to\pm\infty$, and $L$ controls how the grid points are distributed between the nucleus and the far field.
We extend the fields as even functions to $r<0$, which makes them even in $\xi$, and expand them in Chebyshev polynomials,
\begin{align}
    \boldsymbol u(\xi) = \sum_{k=0}^{N-1} \boldsymbol a_k\,T_k(\xi)
    \;,
    \label{S-eqn:cheb_expansion}
\end{align}
where the coefficients $\boldsymbol a_k$ are determined by the values of $\boldsymbol u$ at the $N$ roots $\xi_j=\cos[(2j+1)\pi/(2N)]$ of $T_N$, with $j=0,\dots,N-1$.
For even $N$, these nodes avoid both $\xi=0$ and $\xi=\pm1$, and the even extension ensures regularity at $r=0$.
Derivatives and integrals with respect to $r$ are then evaluated on this expansion using the chain rule.
We use $N=256$ and $L=3$.

Instead of integrating \Eqsref{S-eqn:rq_dyn}--\eqref{S-eqn:gad_dyn} simultaneously, we alternate between two steps.
First, at fixed $\boldsymbol v$, we evolve $\boldsymbol u$ according to \Eqref{S-eqn:gad_dyn} with a semi-implicit scheme that treats the gradient terms implicitly.
Second, we replace $\boldsymbol v$ by the eigenvector of $\boldsymbol H$ with the smallest eigenvalue, computed by an Arnoldi iteration.
We repeat these two steps until three criteria hold simultaneously.
First, the maximal value of $|\partial_t\boldsymbol u|$ from \Eqref{S-eqn:gad_dyn} falls below $5\times10^{-4}$.
Second, the overlap $|\boldsymbol v^\mathrm{old}\cdot\boldsymbol v^\mathrm{new}|$ of successive normalized directions deviates from unity by less than $10^{-4}$.
Third, the far field matches the reservoir to within $10^{-7}$ in $\phi$ and $10^{-6}$ in $\varepsilon$.

Since GAD converges to the index-one saddle point whose basin contains the initial profile, the initial profile determines which nucleus is found.
Where dense and shallow nuclei coexist, we obtain both by starting from dense and from shallow initial profiles, respectively.
All critical nuclei reported in the main text use $\phig=0.9$, $l=20$, $\theta=1$, $\kappa=4$, and $\vartheta=0.5$.

\begin{figure}
    \centering
    \includegraphics[width=0.5\linewidth]{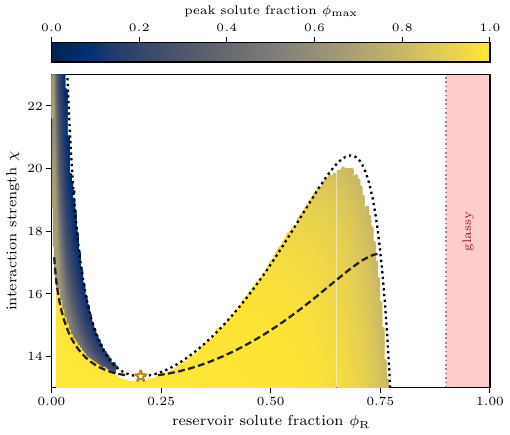}
    \caption{\textbf{The peak solute fraction of critical nuclei jumps across the metastable binodal.}
    Peak solute fraction $\phi_\mathrm{max}$ of the critical nucleus as a function of the reservoir solute fraction $\phiR$ and the interaction strength $\chi$, with lines as in Fig.~3b of the main text.
    The jump coincides with the boundary of the metastable coexisting liquids (dashed line).}
    \label{S-fig:si_phi_max}
\end{figure}

\Figref{S-fig:si_phi_max} shows the peak solute fraction $\phi_\mathrm{max}$ of the critical nuclei over the same range of $\phiR$ and $\chi$ as Fig.~3b of the main text.
Like the excess solute $N_\mathrm{exc}$, it jumps across the boundary of the metastable coexisting liquids.
Dense nuclei reach and even exceed $\phig$, e.g., $\phi_\mathrm{max}=0.999$ at $\phiR=0.05$ and $\chi=13$, while shallow nuclei have $\phi_\mathrm{max}$ well below $\phig$.

\end{document}